\documentclass[prd,a4paper,showpacs,preprint,byrevtex]{revtex4}
\usepackage{graphicx}
\usepackage{dcolumn}
\usepackage{amsmath}
\usepackage{array}
\usepackage{bm}

\begin{document}

\title{Retardation effects in the rotating string model}

\author{Fabien \surname{Buisseret}}
\thanks{FNRS Research Fellow}
\email[E-mail: ]{fabien.buisseret@umh.ac.be}
\author{Claude \surname{Semay}}
\thanks{FNRS Research Associate}
\email[E-mail: ]{claude.semay@umh.ac.be}
\affiliation{Groupe de Physique Nucl\'{e}aire Th\'{e}orique,
Universit\'{e} de Mons-Hainaut,
Acad\'{e}mie universitaire Wallonie-Bruxelles,
Place du Parc 20, BE-7000 Mons, Belgium}

\date{\today}

\begin{abstract}
A new method to study the retardation effects in mesons is presented.
Inspired from the covariant oscillator quark model, it is applied to the
rotating string model in which a non zero value is allowed for the
relative time between the quark and the antiquark. This approach leads
to a retardation term which behaves as a perturbation of the meson mass
operator. It is shown that this term preserves the Regge trajectories
for light mesons, and that a satisfactory agreement with the
experimental data can be obtained if the quark self-energy contribution
is added. The consequences of the retardation on the Coulomb interaction
and the wave function are also analyzed.
\end{abstract}


\eprint{hep-ph/0505168}

\maketitle

\section{Introduction}
\label{sec:intro}

The retardation effect between two interacting particles is a
relativistic phenomenon, due to the finiteness of the interaction speed.
The light mesons are typical systems in which the retardation
mechanisms can significantly contribute to the dynamics, since the light
quark can move at a speed close to the speed of light. To take into
account this effect into effective models, one has to work with a fully
covariant theory. The most elegant approach, from a theoretical point of
view, is the formalism with constraints \cite{sazd86,crat87}, but it
quickly leads to complex equations, uneasy to deal with if one wants to
get analytical or numerical results. Another covariant approach of
mesons is the covariant oscillator quark model (COQM), which allows to
find an analytical expression for the wave function and numerical
results in good agreement with the data \cite{ishi86,ishi92}.
Unfortunately, this model uses a quadratic potential, this form being
different from the linear potential, commonly assumed to describe the
confining part of the interaction and suggested by lattice calculations.
Moreover, vanishing quark masses are not allowed in this approach.

Apart from these two approaches, most of the effective models found in
the literature are based on the equal time ansatz, which simply takes
the time coordinates of both particles to be equal, neglecting the
retardation effects. This procedure allows to deal with simpler
equations, and relativistic corrections can be obtained by developing an
expansion in $v^2/c^2$ of
the model considered, like the Bethe-Salpeter equation or even the QCD
Lagrangian \cite{barc89,olss92}.

The model we propose here is an attempt to include retardation effects
into the rotating string model (RSM) \cite{dubi94,guba94} without making
such an expansion, in order to estimate the retardation contribution
for light quarks. The RSM is an effective model derived from the
QCD Lagrangian, describing a meson by two spinless quarks linked by a
straight string. It has been shown that the RSM was classically
equivalent to the relativistic flux tube model \cite{sema04,buis042}.
This last model, firstly presented in Ref.~\cite{laco89,tf_2}, yields
meson spectra in good agreement with the experimental data
\cite{tf_Semay}. Our method, inspired from the COQM, relies on the
hypothesis that the relative time between the
quark and the antiquark must have a non zero value. In our framework,
the evolution parameter of the system is not the common proper time of
the quarks and the string, but the time coordinate of the center of
mass which plays the role of an ``average" time.

Our paper is organized as follows. In Sec.~\ref{model}, we present the
general formalism of our approach. We compute the retardation
contribution to the meson mass in Sec.~\ref{retard}. We find different
approximations for this contribution in Sec.~\ref{approx}, with a
special interest for light quarks, and we numerically study the
retardation effects in Sec.~\ref{num}. As our model relies on unusual
hypothesis, it is worth comparing our results with those of other
existing models. This is done in Sec.~\ref{compar}. Finally, we compare
the meson spectra of our RSM including the retardation term with the
experimental data in Sec.~\ref{data}. Some concluding remarks are given
in Sec.~\ref{conclu}. The appendix contains some useful formulas.

\section{The rotating string model with non zero relative time}
\label{model}

It has been shown that, starting from the QCD Lagrangian and neglecting
the spin contribution of the quark and the antiquark, the Lagrange
function of a meson can be built from a Nambu-Goto action \cite{dubi94}
which reads ($\eta={\rm diag}(+---)$ and $\hbar=c=1$)
\begin{equation}
\label{nambu}
{\cal L}(\tau)=-m_{1}\sqrt{\dot{\bm x}^{2}_{1}}-m_{2}\sqrt{\dot{\bm
x}^{2}_{2}}
-a\int^{1}_{0}d\beta\sqrt{(\dot{\bm w}\bm w')^{2}-\dot{\bm w}^{2}{\bm
w'}^{2}}.
\end{equation}
The two first terms are the kinetic energy operators of the quark and
the antiquark, whose current masses are $m_{1}$ and $m_{2}$. These two
particles are attached by a string with a tension $a$. $\bm x_{i}$ is
the coordinate of the quark $i$ and $\bm w$ is the coordinate of the
string.
$\bm w$ depends on two variables defined on the string worldsheet: One
is spacelike, $\beta$, and the other timelike, $\tau$. Derivatives are
denoted $\bm w'=\partial_{\beta}\bm w$ and
$\dot{\bm w}=\partial_{\tau} \bm w$. In this picture, $\tau$ is a common
proper time for the string and the quarks. Introducing auxiliary fields
to get
rid of the square root in the Lagrangian~(\ref{nambu}) and making the
straight line ansatz to describe the string, an effective Lagrangian can
be derived \cite{guba94}
\begin{eqnarray}
\label{step2}
{\cal L}&=& -\frac{1}{2} \left[
\frac{m^{2}_{1}}{\mu_{1}}+\frac{m^{2}_{2}}{\mu_{2}}+
a_{1}
\dot{\bm R}^{2}+2a_{2}\dot{\bm R}\dot{\bm r}-2(c_{1}+\dot{\zeta}a_{1})
\dot{\bm R}\bm r\right.\nonumber\\
&&\left.-2( c_{2}+\dot{\zeta}a_{2})\dot{\bm r}\bm r
+a_{3}\dot{\bm r}^{2}+(a_{4}+2\dot{\zeta}c_{1}+\dot{
\zeta}^{2}a_{1})\bm r^{2}
\right],
\end{eqnarray}
where the coefficients $a_{1}$, $a_{2}$, \dots, given in the
appendix, depend on various auxiliary fields $\mu_{1}$, $\mu_{2}$,
$\nu$, and $\eta$. The parameter $\zeta$ defines the position $\bm R$ of
the
center of mass: $\bm R=\zeta \bm x_1+(1-\zeta) \bm x_2$. $\bm r$ is the
relative coordinate $\bm r=\bm x_{1}-\bm x_{2}$. The auxiliary fields
$\mu_{1}$ and $\mu_{2}$ are seen as constituent masses for the quarks,
and
$\nu$ can be interpreted in the same way as an effective energy for the
string whose ``static" energy is $a r$ \cite{sema04,buis042}. Let us
note that the straight line ansatz for the string implies
$\bm w=\bm R+(\beta-\zeta)\bm r$.

The total and relative momentum, computed from the
Lagrangian~(\ref{step2}), are respectively
\begin{subequations}
\label{momentum}
\begin{eqnarray}
P_{\mu}&=&\frac{\partial {\cal L}}{\partial \dot{R}^{\mu}}=
-a_{1}\dot{R}
_{\mu}-a_{2} \dot{r}_{\mu}+(c_{1}+\dot{\zeta}a_{1}) r_{\mu}, \\
\label{prel}
p_{\mu}&=&\frac{\partial {\cal L}}{\partial \dot{r}^{\mu}}=
-a_{2}\dot{R}
_{\mu}-a_{3} \dot{r}_{\mu}+(c_{2}+\dot{\zeta}a_{2}) r_{\mu}.
\end{eqnarray}
\end{subequations}
As we will work in the center of mass frame, the total vector momentum
$\vec{P}$ of the system must vanish, which implies that
\begin{equation}
\label{cm_rep}
\dot{\vec{R}}=\frac{(c_{1}+\dot{\zeta}a_{1})\vec{r}-a_{2}\dot{\vec{r}}}
{a_{1}}.
\end{equation}
Moreover, the relative vector momentum $\vec{p}$ is given by
\begin{equation}
\label{p_rel}
\vec{p}=a_{2}\dot{\vec{R}}+a_{3}\dot{\vec{r}}-(c_{2}+\dot{\zeta}a_{2})
\vec{r}.
\end{equation}
Thus, we impose $a_2=0$ in order that $\vec{p}$ does not depend on the
motion of the center of mass. This leads to the following value for
the parameter $\zeta$
\begin{equation}
\label{zeta}
\zeta=\frac{\mu_{1}+\int^{1}_{0} d\beta\, \beta\,
\nu}{\mu_{1}+\mu_{2}+\int^{1}_{0} d\beta\, \nu}.
\end{equation}
Equation~(\ref{zeta}) reduces to $\zeta=1/2$ in the symmetrical case
($m_{1}=m_{2}$) and to $\zeta=m_{1}/(m_{1}+m_{2})$ in the
nonrelativistic limit, as expected \cite{buis042}.

To go a step further, one usually takes the temporal coordinates of
the quarks and the string to be equal to the common proper time $\tau$,
this time being also the time $t$ in the center of mass frame
\begin{equation}\label{tegal}
x^{0}_{1}=x^{0}_{2}=w^{0}=\tau=t.
\end{equation}
Then, we have $\bm r=(0,\vec{r}\, )$, $\bm R=(t,\vec{R}\ )$,
$\dot{\bm r}=(0,\dot{\vec{r}}\, )$, and
$\dot{\bm R}=(1,\dot{\vec{R}}\, )$. This procedure allows to deal with
simpler equations, but neglects the relativistic retardation
effects due to a possible non zero value of the relative time $r^{0}$.
Since
these effects are precisely those we want to study in this paper, we
have to make a less restrictive hypothesis. As in the formalism of the
COQM \cite{ishi86,ishi92}, we define
\begin{equation}\label{choice1}
\bm r=(\sigma,\vec{r}\,),\ \bm R=(\bar{t},\vec{R}\, ).
\end{equation}
The temporal coordinate of the center of mass, $\bar{t}$, can be seen as
an ``average time" for the meson. This is particularly clear in
the symmetrical case, where $\bar{t}=(x^{0}_{1}+x^{0}_{2})/2$. Our
choice is to take $\bar{t}$ as the evolution parameter of the system. We
identify it as the common proper time for the quarks and the string, and
the dotted quantities are derived with respect to $\bar{t}$.
We have for example
\begin{equation}\label{choice2}
\dot{\bm r}=(\dot{\sigma},\dot{\vec{r}}\, ),\ \dot{\bm
R}=(1,\dot{\vec{R}}\, ).
\end{equation}
The special case $\sigma=0$ is equivalent to the relation~(\ref{tegal}).

The Lagrangian (\ref{step2}) can now be rewritten using
formulas~(\ref{cm_rep}), (\ref{p_rel}), (\ref{choice1}), and
(\ref{choice2}) as
\begin{equation}
{\cal L}={\cal L}_{0}+\Delta {\cal L},
\end{equation}
with
\begin{eqnarray}
\label{lagr_1}
{\cal L}_{0}&=& -\frac{1}{2} \left[
\frac{m^{2}_{1}}{\mu_{1}}+\frac{m^{2}_{2}}{\mu_{2}}+a_{1}+
\frac{1}{a_{1}}\left((c^{2}_{1}-a_{4}a_{1})\vec{r}^{\, 2}-a_{3}
a_{1}\dot
{\vec{r}}^{\, 2}+2a_{1}c_{2}\dot{\vec{r}}\, \vec{r}\right) \right],
\\
\label{deltaL}
\Delta{\cal L}&=&(c_{1}+\dot{\zeta}a_{1})\sigma+c_{2}\dot{\sigma}\sigma-
\frac{a_{3}}{2}\dot{\sigma}^{2}-\frac{1}{2}(a_{4}+2\dot{\zeta}c_{1}+\dot
{\zeta}^{2}a_{1})\sigma^{2}.
\end{eqnarray}
We have gathered the relative time dependent terms in $\Delta{\cal L}$,
which contains the contribution of the retardation. Let us remark that
the Lagrangian ${\cal L}_{0}$ does not depend on $\dot{\zeta}$.

We shall consider in the following $\Delta{\cal L}$ as a
perturbation of ${\cal L}_{0}$. With this hypothesis, the auxiliary
fields can be
eliminated by considering only the constraint $\delta{\cal L}_{0}=0$.
In Ref.~\cite{buis042}, it is shown how a set of three equations can
be derived from the Lagrangian~(\ref{lagr_1}), to define the usual
rotating string model (RSM)
\begin{subequations}
\label{seteq1}
\begin{eqnarray}
0&=&\mu_{1}y_{1}-\mu_{2}y_{2}-\frac{ar}{y_t}\left(\sqrt{1-y^{2}_{1}}-
\sqrt{1-y^{2}_{2}}\right), \label{peq0} \\
\frac{L}{r}&=&\frac{1}{y_t}(\mu_{1}y^{2}_{1}+\mu_{2}y^{2}_{2})+\frac{
ar} {y_t^{2}} (F(y_{1})+F(y_{2})), \label{lor} \\
H_{0}&=&\frac{1}{2}\left[\frac{p^{2}_{r}+m^{2}_{1}}{\mu_{1}}+\frac{p^{2}
_{r} +m^{2}_{2}}{\mu_{2}}+\mu_{1}(1+y^{2}_{1})+\mu_{2}(1+y^{2}_{2})
\right]
\nonumber \\
&&+\frac{ar}{y_t}(\arcsin y_{1}+\arcsin y_{2}), \label{haux}
\end{eqnarray}
with
\begin{equation}
\label{fy}
F(y_{i})=\frac{1}{2}\left[\arcsin y_{i}-y_{i}\sqrt{1-y^{2}_{i}}
\right] \quad \text{and} \quad y_t=y_{1}+y_{2}.
\end{equation}
\end{subequations}
$p_{r}$ is the radial momentum and $y_{i}$ can be seen as the transverse
velocities of the quark $i$. The first relation gives the cancellation
of the
total momentum in the center of mass frame, while the two last ones
define
respectively the angular momentum and the Hamiltonian. As we can see
in Eqs.~(\ref{seteq1}), the only remaining auxiliary fields are
$\mu_{i}$. The extremal values of the auxiliary fields $\eta$ and $\nu$
are given by \cite{buis042}
\begin{subequations}
\begin{eqnarray}\label{etadef}
\eta_{0}&=&\kappa\left(\beta-\phi\right),
\\ \label{nu0}
\nu_{0}&=&\frac{ar}{\sqrt{1-y_t^{2}(\beta-\zeta)^{2}}},
\end{eqnarray}
\end{subequations}
with
$\kappa=-\frac{\vec p\cdot \vec r}{\tilde{\mu} r^2}$,
$\phi=\frac{\mu_{1}}{\mu_{1}+\mu_{2}}$, and
$\tilde{\mu}=\frac{\mu_{1}\mu_{2}}{\mu_{1}+\mu_{2}}$.
Let us remark that a closed form can not be obtained for the Hamiltonian
$H_0$ because it impossible to eliminate analytically the variables
$y_1$ and $y_2$ as a
function of $L$, by means of the two first Eqs.~(\ref{seteq1}).

Since we have a contribution from the relative time, the total
Hamiltonian is given by
\begin{equation}
H=H_{0}+\Delta H,
\end{equation}
with
\begin{equation}
\Delta H=\Sigma \dot{\sigma}-\Delta {\cal L}.
\end{equation}
With $\Sigma=\partial {\cal L}/\partial \sigma$, Eq.~($\ref{prel}$)
leads to
\begin{equation}
\dot{\sigma}=\frac{c_{2}\sigma-\Sigma}{a_{3}},
\end{equation}
and finally we obtain
\begin{equation}\label{deltaH}
  \Delta H=-\frac{\Sigma^{2}}{2a_{3}}+\frac{c_{2}}{a_{3}}\Sigma \sigma-(
  c_{1}+\dot{\zeta}a_{1}) \sigma -\frac{c_{2}^{2}}{2a_{3}}\sigma^{2}+
  \frac{1}{2}(a_{4}+2\dot{\zeta}c_{1}+\dot{\zeta}^{2}a_{1})\sigma^{2},
\end{equation}
the perturbation of the RSM Hamiltonian due to the retardation effect.

In the following, for simplicity, we
will focus on the symmetrical case. Then $\dot{\zeta}=0$, and the
Hamiltonian~(\ref{deltaH}) becomes
\begin{equation}\label{deltaH2}
  \Delta H=-\frac{1}{2a_{3}}\left[\Sigma^{2}-2c_{2}\Sigma \sigma +(c_{2}
  ^{2}-a_{4}a_{3})\sigma^{2}\right].
\end{equation}
Wether $\Delta H$ is really a perturbation or not has to be verified. We
will check this hypothesis a posteriori by a numerical computation of
the retardation contribution to the meson masses (see Sec.~\ref{num}).

\section{Retardation effects}
\label{retard}

\subsection{Contribution to the mass}

Up to now, we were working in a classical framework. But in order
to study the influence of the retardation on the meson spectrum, we have
to consider a quantized version of the total RSM Hamiltonian
$H_{0}+\Delta H$. We can thus replace $L$ by $\sqrt{\ell(\ell+1)}$ and
consider $r$ and $\sigma$ as operators such that
\begin{equation}
  \left[r,\, p_{r}\right]=i,\ \left[\sigma,\, \Sigma\right]=-i.
\end{equation}
\par The total Hamiltonian has schematically the following structure
\begin{equation}
H(\sigma,r)=H_{0}(r)+\Delta{H}(\sigma,r).
\end{equation}
The relative time $\sigma$ only appears in the perturbation, and
$H_{0}$ only depends on the radius $r$. So, we make the following ansatz
to write the total wave function
\begin{equation}
\left|\psi (\bm r)\right\rangle=\left|R(\vec r\,)\right\rangle\otimes
\left |A(\sigma)\right\rangle,
\end{equation}
where $\left|R(\vec r\,)\right\rangle$ is a solution of the
eigenequation
\begin{equation}\label{eigen1}
H_{0}(r)  \left|R(\vec r\,)\right\rangle=M_{0}\left|R(\vec
r\,)\right\rangle.
\end{equation}
Such a problem can be solved, for instance, by the Lagrange mesh
technique \cite{buis041}. The total mass is written
\begin{eqnarray}
M&=&M_{0}+\left\langle  A(\sigma) \right| \otimes\left\langle  R(\vec r
\,) \right| \Delta H(r,\sigma)\left|R(\vec r\,)\right\rangle\otimes\left
|A(\sigma) \right\rangle \nonumber\\
&=&M_{0}+\Delta M. \label{mtot}
\end{eqnarray}
the contribution $\Delta M$ is then given by the solution of the
eigenequation
\begin{equation}\label{eigenret}
\Delta {\cal H}(\sigma)\left|A(\sigma)\right\rangle=\Delta M  \left|A(
\sigma)\right\rangle,
\end{equation}
where
\begin{equation}\label{average}
\Delta{\cal H}(\sigma)= \left\langle R(\vec r\,) \right| \Delta
H(r,\sigma) \left|R(\vec r\,)\right\rangle.
\end{equation}
In order to eliminate the unphysical excitations of the relative time,
we consider only the ground state of the Hamiltonian
$\Delta{\cal H}(\sigma)$, as it is done in Refs.~\cite{ishi86,ishi92}.
Using formula~(\ref{deltaH2}), this Hamiltonian is
defined by
\begin{equation}\label{dH3}
\Delta {\cal H}\approx-\frac{1}{2\langle
a_{3}\rangle}\left[\Sigma^{2}-\langle c_{2}\rangle\{\Sigma, \sigma\} +
\langle c_{2}^{2}-a_{4}a_{3}\rangle\sigma^{2}\right],
\end{equation}
where all mean values $\langle\rangle$ are computed with a space
function $R(\vec r\,)$ and where $\{ A,B\}=AB + BA$.
We also use the approximation
$\langle 1/x\rangle\approx1/\langle x\rangle$.

Using Eqs.~(\ref{etadef}) and (\ref{acdef3}), we see that
$c_{2} \propto \kappa$.
Since we are in the symmetrical case, we can assume
$\langle \kappa \rangle=0$, and so
$\left\langle c_{2}\right\rangle=0$. On the contrary,
$\left\langle c^{\, 2}_{2}\right\rangle\not=0$ because
$\left\langle p^{\, 2}_{r}\right\rangle>0$. Finally the Hamiltonian
$\Delta{\cal H}$ takes its final form
\begin{equation}\label{dH4}
\Delta {\cal H}=-\frac{1}{2\langle a_{3}\rangle}\left[\Sigma^{2}+
\langle c_{2}^{2}-a_{4}a_{3}\rangle\sigma^{2}\right].
\end{equation}
With our approximations, the retardation contribution to the Hamiltonian
looks like an harmonic oscillator for the canonical variables
$(\sigma,\Sigma)$. Let us define
\begin{equation}
\rho^{2}= c_{2}^{2}-a_{4}a_{3}.
\end{equation}
Thanks to formulas~(\ref{acdef}), we can compute $\rho^{2}$
\begin{eqnarray}\label{rho2}
\rho^{2}&=&\frac{a}{2ry}\left(\frac{\mu}{2}+\frac{ar}{8y^{3}}\left(-y
\sqrt{1-y^{2}}+\arcsin y\right)\right)\left(y\sqrt{1-y^{2}}+\arcsin y
\right)\nonumber\\
&-&\frac{ap^{2}_{r}}{4\mu r y^{3}}\left(-y\sqrt{1-y^{2}}+\arcsin
y\right).
\end{eqnarray}
Assuming that $\left\langle \rho^{2}\right\rangle>0$ (this is
checked in Sec.~\ref{approx}), the ground state solution of the
eigenequation~(\ref{eigenret}) is given by
\begin{subequations}
\label{dMetc}
\begin{eqnarray}\label{dM}
\Delta M&=&-\frac{1}{2}\omega ,
\\\label{fo_ret}
A(\sigma)&=&\left(\frac{\beta}{\pi}\right)^{1/4}\exp\left(-\frac{\beta}{
2
}\sigma^{2}\right),
\end{eqnarray}
with
\begin{eqnarray}\label{betadef}
\beta&=&\sqrt{\left\langle \rho^{2}\right\rangle},
\\\label{omegadef}
\omega&=&\frac{\beta}{\left\langle a_{3}\right\rangle}.
\end{eqnarray}
\end{subequations}
An immediate conclusion to draw from Eqs.~(\ref{dMetc}) is
that the retardation effects bring a negative contribution to the meson
masses, and that the more probable value for the relative time is
$\sigma=0$.
It is worth noting that these results
are formally identical to those of the COQM. In
Sec.~\ref{compar}, we compare the two approaches with more details.

\subsection{Modification of the coulomb term}

In the RSM, the string contribution takes into account the interactions
at large distances, which are responsible for the confinement. To make
more realistic models, it is necessary to add short range potentials
\cite{tf_Semay}. For instance, the one gluon exchange mechanism can be
simulated by a Coulomb term
\begin{equation}
V_{C}(r)=-\frac{4}{3}\frac{\alpha_{S}}{r},
\end{equation}
with $\alpha_{S}$ the strong coupling constant. This formula must be
modified if we consider the retardation effects. Indeed, the
quark and the antiquark are able to exchange one gluon if their
separation $\bm r$ is light-like, that is to say if
$\sigma^{2}-r^{2}=0$. The probability for $\sigma$ to
be negative or positive is
\begin{equation}
p\ (\sigma<0)=\int^{0}_{-\infty}d\sigma A(\sigma)^{2}=
\frac{1}{2}=p\, (\sigma>0).
\end{equation}
Consequently, as in the COQM
\cite{ishi86,ishi92}, we make the following substitution
\begin{equation}
V_{C} \rightarrow V_{C}\frac{1}{2\lambda}\left[ \delta(\sigma+r)+\delta(
\sigma-r)\right],
\end{equation}
where $\lambda$ is an energy scale which is introduced so as to give the
correct dimension. This parameter is purely phenomenological and could
depend on the quark masses \cite{ishi86}. In this paper, we assume that
$\lambda$ is a constant.
The effective Coulomb potential $\tilde{V}_{C}$, treated as a
perturbation, is then computed with
the relation
\begin{equation}
\tilde{V}_{C}=\int^{+\infty}_{-\infty}d\sigma A(\sigma)^{2}
V_{C}\frac{1}{2\lambda}\left[ \delta(\sigma+r)+\delta(\sigma-r)\right],
\end{equation}
and we obtain a damped effective Coulomb potential,
\begin{equation}\label{coul_eff}
\tilde{V}_{C}=-\frac{4}{3}\frac{\alpha_{S}}{\lambda
r}\left(\frac{\beta}{\pi}
\right)^{1/2} \exp\left(-\beta r^{2}\right).
\end{equation}

\section{Approximations for $\Delta M$}
\label{approx}

The key ingredient to compute the retardation term (\ref{dM}) is the
knowledge of $\rho^{2}$ and $a_{3}$, respectively given by
formulas~(\ref{rho2}) and (\ref{acdef3}). As these expressions are
complicated, we will try to get various simpler ones following the value
of the quark mass $m$.
In the following we will use two limiting cases for
$a_3$ and $\rho^2$ quantities: $y=0$ corresponds to a vanishing angular
momentum or to a very high mass, and $y=1$ correspond to a very high
angular momentum or to a very small mass. Both situations are summarized
in Table~\ref{tab:a3r}.
\begin{table}[htb]
\caption{Values of $a_3$ and $\rho^2$ for $y=0$ and $y=1$.}
\label{tab:a3r}
\begin{ruledtabular}
\begin{tabular}{ccc}
 & $y=0$ & $y=1$ \\
\hline
$a_3$ & $\dfrac{\mu}{2}+\dfrac{a r}{12}$ &
$\dfrac{\mu}{2}+\dfrac{\pi a r}{16}$\\
$\rho^2$ &
$\dfrac{a \mu}{2 r}+\dfrac{a^2}{12} -\dfrac{a p^{2}_{r}}{6 \mu  r}$ &
$\dfrac{\pi a \mu}{8 r}+\dfrac{\pi^2 a^2}{64} -\dfrac{\pi a p^{2}_{r}}{8
\mu r}$ \\
\end{tabular}
\end{ruledtabular}
\end{table}

The solutions of the RSM equations are, in good approximation, very
close to the solutions of a two-body spinless Salpeter equation with a
linear potential
\begin{equation}
\label{muss}
H^{SS}=2 \sqrt{\vec p\,^2+m^{2}} + a r
\end{equation}
and with the pure string correction treated as a perturbation
\cite{bada02}.
For massless quarks, this correction is only about 6\% of the
meson mass \cite{buis043}. So, we will work with the
Hamiltonian $H^{SS}$ without the string correction, except in
Sec.~\ref{data}. Actually,
this approximation is equivalent to consider the RSM at the order
$y^2$ \cite{buis043}. Within this framework, the extremal value of the
auxiliary field $\mu$ is \cite{sema04,buis043}
\begin{equation}
\label{mu}
\mu=\sqrt{\vec p\,^2+m^{2}}.
\end{equation}
We have then
\begin{equation}
\label{muss2}
\langle p^{2}_{r}\rangle \approx \langle \mu
\rangle^{2}-m^{2}-\frac{\ell(\ell+1)}{\langle r \rangle^{2}},
\end{equation}
and $0 < \langle p^{2}_{r}\rangle \le \langle \mu \rangle^{2}$.

\subsection{High quark mass}

If we assume that $\langle \mu\rangle \gg a\langle r\rangle$, we can set
$y \approx 0$. In this case, the relations~(\ref{dM}) and
(\ref{betadef}) with $\langle \mu\rangle \gg a\langle r\rangle$ reduces
to
\begin{eqnarray}
\beta_{h}&\approx&\sqrt{\frac{a\langle \mu\rangle}{2
\langle r\rangle}}\sqrt{1-\frac{\langle p^{2}_{r}
\rangle}{3\langle \mu\rangle^{2}}},
\\
\Delta M_{h}&\approx&-\sqrt{\frac{a}{2\langle \mu\rangle
\langle r\rangle}}\sqrt{1-\frac{\langle p^{2}_{r}
\rangle}{3\langle \mu\rangle^{2}}}.
\end{eqnarray}
It is clear that $\langle \rho^2\rangle$ is positive.
If $m$ is very large, the dynamical effects can be neglected and, using
$\langle\mu \rangle\approx m$, we have
\begin{eqnarray}\label{betahh}
\beta_{hh}&\approx&\sqrt{\frac{a m}{2\langle r\rangle}},
\\ \label{omegahh}
\Delta M_{hh}&\approx&-\sqrt{\frac{a}{2m\langle r\rangle}}.
\end{eqnarray}
The eigenvalues $M_0$ are then given with a good accuracy by
\cite{sema92}
\begin{equation}
M_0 \approx 2m+ \left(\frac{a^{2}}{m}\right)^{1/3}\epsilon_{n\ell},
\end{equation}
where $\epsilon_{n\ell}$ is an eigenvalue of the dimensionless
Hamiltonian $\left(\vec{q}\,^2+|\vec x\,|\right)$.
The nonrelativistic virial theorem implies that
\begin{equation}
a\langle r\rangle=\frac{2}{3}(M_0-2m)=\frac{2}{3}\left(
\frac{a^{2}}{m}\right)^{1/3}\epsilon_{n\ell}.
\end{equation}
So we obtain
\begin{eqnarray}\label{betahh2}
\beta_{hh}&\approx&\frac{(a m)^{2/3}}{2}
\sqrt{\frac{3}{\epsilon_{n\ell}}},
\\ \label{omegahh2}
\Delta M_{hh}&\approx&-\frac{1}{2}\left(\frac{a^2}{m}\right)^{1/3}
\sqrt{\frac{3}{\epsilon_{n\ell}}}.
\end{eqnarray}
Approximate values for the quantities $\epsilon_{n\ell}$ can be found in
Ref.~\cite{buis043}.

\subsection{Vanishing quark mass}

In this section, we will work with the
Hamiltonian $H^{SS}$ for $m=0$. With these conditions, the relativistic
virial theorem \cite{luch90} gives the following results \cite{sema04}
\begin{eqnarray}\label{rvt1}
M_0 \approx \langle H^{SS}\rangle &=& 4\langle \mu\rangle, \\
\label{rvt2}
a\langle r\rangle &=& 2\langle \mu\rangle.
\end{eqnarray}

Let us first focus on the case $\ell=0$, for which $y=0$ and
$\langle p_r^2 \rangle \approx \langle \mu\rangle^2$.
Thanks to these relations, we have for light quarks
\begin{eqnarray}
\label{betal}
\left.\beta_{l}\right|_{\ell=0}&=&\frac{a}{2},
\\ \label{omegal}
\left.\Delta M_{l}\right|_{\ell=0}&=&-\frac{3a}{2M_0}.
\end{eqnarray}
Secondly, let us consider the case $\ell \gg 1$, for which we can assume
that $y=1$. We then find
\begin{eqnarray}
\left.\beta_{l}\right|_{\ell\gg 1}&=&\frac{a}{4}\sqrt{\pi\left(1+\frac{
\pi
}{4}-\frac{16 \langle p^{2}_{r}\rangle}{ M_0^{2}}\right)},
\\ \label{omegal2}
\left.\Delta M_{l}\right|_{\ell\gg 1}&=&-\frac{a}{M_0}\left(\frac{4}{4+
\pi}\right)\sqrt{\pi\left(1+\frac{\pi}{4}-\frac
{16 \langle p^{2}_{r}\rangle}{M_0^{2}}
\right)}.
\end{eqnarray}
Combining Eqs.~(\ref{muss2}), (\ref{rvt1}), and (\ref{rvt2}), we find
\begin{equation}
\label{promu}
\langle p^{2}_{r}\rangle \approx \langle \mu\rangle^{2}-
\frac{64 a^2\ell(\ell+1)}{M_0^{4}}.
\end{equation}
Consequently, for massless quarks with high angular momentum, we have
\begin{eqnarray}
\left.\beta_{l}\right|_{\ell\gg 1}&=&\frac{a}{4}\sqrt{\pi\left(\frac{
\pi}{4}+\frac{64 a^2\ell(\ell+1)}{M_0^4}\right)},
\\
\left.\Delta M_{l}\right|_{\ell\gg 1}&=&-\frac{a}{M_0}\left(\frac{4}{4+
\pi}\right)\sqrt{\pi\left(\frac{
\pi}{4}+\frac{64 a^2\ell(\ell+1)}{M_0^4}\right)}.
\end{eqnarray}
Again, $\langle \rho^2\rangle$ is positive for $\ell=0$ and $\ell \gg
1$.

In good approximation, it appears that
\begin{equation}\label{propo}
\Delta M_{l}\propto\frac{1}{M_{0}}.
\end{equation}
The square meson mass composed of light quarks is then given by
\begin{equation}
\label{m2meson}
M^{2}_{l}\approx M^{2}_{0}+2M_{0}\, \Delta M_{l},
\end{equation}
where the term $\Delta M_{l}^2$ is neglected.
This shows that the retardation term only causes a global shift
of the square meson masses and consequently preserves the Regge
trajectories, since
$M^{2}_{0}\propto\ell$ for large values of $\ell$ \cite{bada02,buis043}.

\section{Numerical results}
\label{num}

\begin{figure}[htb]
\includegraphics*[width=8.0cm]{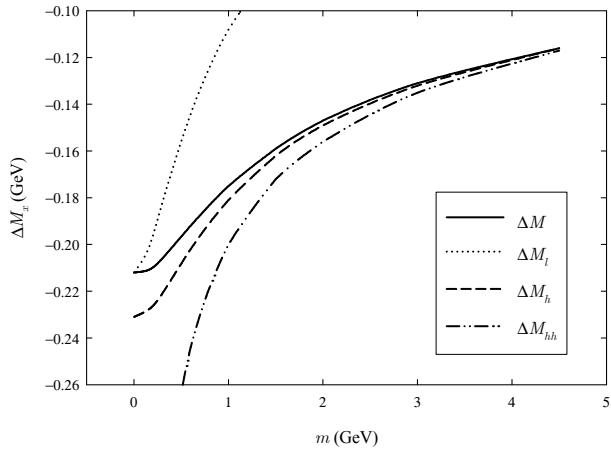}
\caption{Comparison between the exact retardation term and the
different approximations of Sec.~\ref{approx} for various quark
masses, with $a=0.2$~GeV$^{2}$ and $\ell=n=0$. }
\label{fig:dM_m}
\end{figure}

In order to obtain better values for the contribution of the
retardation, we will compute it at the second order in $y$. In this
case, the quantities $a_3$ and $\rho^2$ become
\begin{subequations}
\label{a3rho2}
\begin{eqnarray}
\label{a3num}
\langle a_{3}\rangle &\approx& \left(\frac{a\langle r
\rangle}{12}+\frac{\langle
\mu\rangle}{2}\right)+\frac{1}{40}a\langle r\rangle
\langle y^{2}\rangle, \\\label{rho2num}
\langle \rho^{2}\rangle &\approx& \left(\frac{a^{2}}{12}-\frac{
a\langle p^{2}_{r}\rangle}{6\langle \mu
\rangle\langle  r\rangle}+\frac{a\langle \mu
\rangle}{2\langle r\rangle}\right)+\left(\frac{a^{2}}{90}-
\frac{a \langle p^{2}_{r}\rangle}{20\langle \mu
\rangle\langle  r\rangle}-\frac{a\langle \mu
\rangle}{12 \langle r\rangle}\right)\langle
y^{2}\rangle.
\end{eqnarray}
These expressions can be calculated using the relation \cite{buis043}
\begin{equation}
\langle y^{2}\rangle \approx \frac{\ell(\ell+1)}{\langle r
\rangle^{2}\left(a\langle r\rangle/6+
\langle \mu\rangle\right)^{2}}.
\end{equation}
\end{subequations}
In the following, the retardation contribution obtained thanks to
Eqs.~(\ref{a3rho2}) will be called ``exact" by opposition to
the more approximate formulas obtained in Sec.~\ref{approx}, and it
will be simply
denoted by $\Delta M$. The exact contribution is compared to the
approximate ones in Fig.~\ref{fig:dM_m} ($\langle \rho^2 \rangle$ is
always positive). In
this graph, $\ell=n=0$, but
the qualitative features of the curves are
the same for other quantum numbers. We can thus determine a validity
domain for each
approximation. In fact, $\Delta M_{l}$ is the best for $m<0.175$~GeV
($u$, $d$ quarks, which are commonly denoted $n$). For
$0.175$~GeV $<m<4.0$~GeV ($s$, $c$ quarks), $\Delta M_{h}$ is rather
good, and for masses larger than $4.0$~GeV ($b$ quark), $\Delta M_{hh}$
works well. As expected, the retardation contribution is less important
when the quark mass is larger, for which a nonrelativistic treatment is
more justified. One could ask how the systematic substitution
$\langle \mu^{2}\rangle\rightarrow\langle
\mu\rangle ^{2}$ does affect the results. Actually, the values
obtained by the two methods differ at most by $5$\%. So, we will
maintain our choice, which is to use systematically
powers of $\langle \mu\rangle$.

\begin{figure}[htb]
\includegraphics*[width=8.0cm]{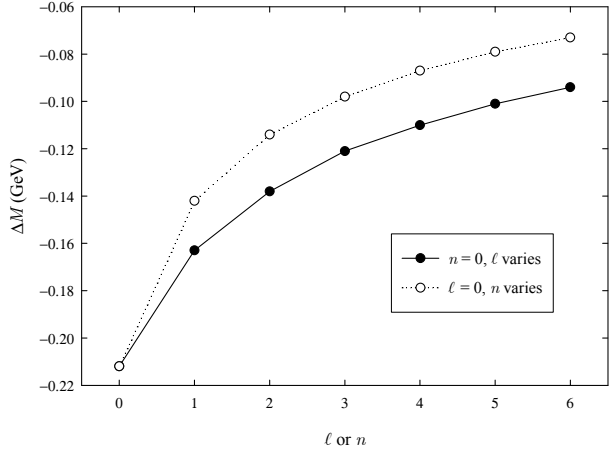}
\caption{Exact retardation term versus: $\ell$ for $n=0$ (filled
circles) and $n$ for $\ell=0$ (empty circles), with $a=0.2$~GeV$^{2}$
and $m=0$.}
\label{fig:dM_ln}
\end{figure}

We focus now on the light quarks, especially the massless case:
The retardation effects are indeed expected to be the largest
when $m=0$, for which the motion is the most relativistic. Typical
behaviors of $\Delta M$
with $\ell$ and $n$ are showed in Fig.~\ref{fig:dM_ln}. If we take
$a=0.2$~GeV$^2$, the ground state mass $M_{0}$ is $1.413$~GeV. The
contribution of the retardation is $-0.205$~GeV. So, in the
worst case, the contribution is about $15$\% of the non perturbed mass.
This result justifies a posteriori the perturbative theory we built in
Secs.~\ref{model} and \ref{retard}. Moreover, we see that, for a fixed
quark mass, the retardation contribution decreases when the
different quantum numbers, $\ell$ or $n$, increase. This means that the
key element is not the quark mass $m$, but the its constituent mass
$\langle \mu \rangle$ \cite{morg99} which also increases with these
quantum numbers. The more the constituent mass is large, the more the
retardation effect is small.

\begin{figure}[htb]
\includegraphics*[width=8.0cm]{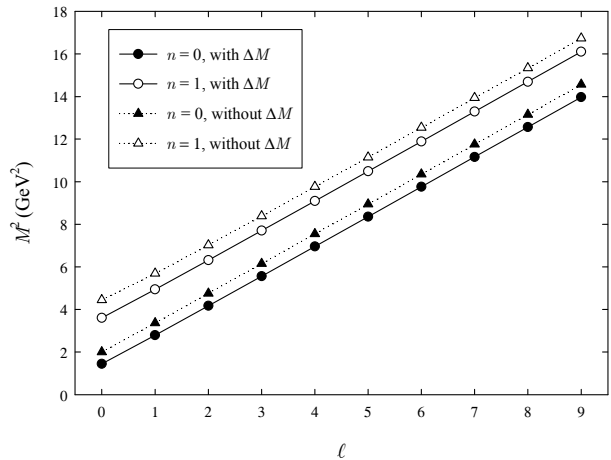}
\caption{Regge trajectories with exact retardation term (circles) and
without (triangles), for $n=0$ and $1$, with $m=0$ and
$a=0.2$~GeV$^{2}$.}
\label{fig:regge}
\end{figure}

In Sec.~\ref{approx}, we showed that the retardation only causes a
global shift of the Regge trajectories for light quarks. Even if this
result is only approximate, we can check in Fig.~\ref{fig:dM_m} that we
can have confidence in our formula $\Delta M_{l}$ when $m=0$. As a
supplementary test in this case, we have plotted in Fig.~\ref{fig:regge}
the square meson masses versus the angular momentum, with and without
the exact retardation term. We clearly see that the linearity of the
Regge trajectories is preserved as well as the slope.

Another interesting quantity to study is $\beta$, given by
Eq.~(\ref{betadef}), the size of the relative time part of the wave
function~(\ref{fo_ret}) and the range of the effective Coulomb
potential~(\ref{coul_eff}).
Let us see what happens for two extreme cases: the $n$ quark and the
$b$ quark, for which we take $m_{n}=0$ and $m_{b}=4.660$~GeV. For
$n=\ell=0$, $\beta_{l}$, used for the $n$ quark, is given by
formula~(\ref{betal}). For the $b$ quark, we can use
formula~(\ref{betahh2}). For a standard value $a=0.2$~GeV$^{2}$, we find
\begin{equation}\label{betaapp}
\left.\beta_{hh}\right|_{\ell=n=0}=0.542~{\rm GeV}^{2}\gg\left.\beta_{l}
\right|_{\ell=n=0}=0.1~{\rm GeV}^{2}.
\end{equation}
So, the wave function is much more peaked around $\sigma=0$ when the
mass increases. The numerical evaluation with Eqs.~(\ref{a3rho2}) gives
\begin{equation}
\left.\beta\right|_{\ell=n=0}=0.533~{\rm GeV}^{2}\gg\left.\beta\right|_{
\ell=n=0}=0.096~{\rm GeV}^{2},
\end{equation}
result very close to the approximate one~(\ref{betaapp}).

\begin{figure}[htb]
\includegraphics*[width=8.0cm]{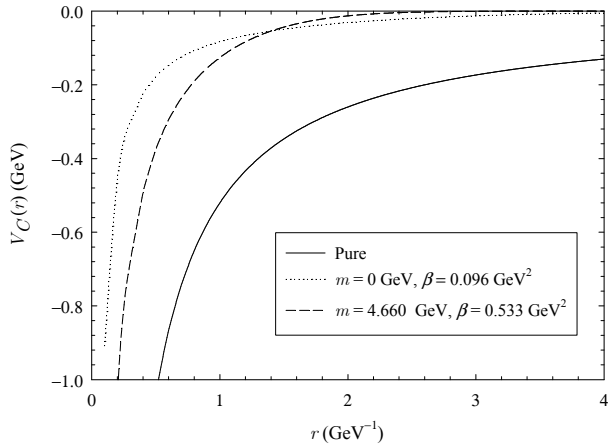}
\caption{Pure Coulomb potential with $\alpha_{S}=0.4$ (straight line),
and effective Coulomb potentials for $m=0$ and $m=4.660$~GeV with
$\lambda=1$~GeV (dotted lines).}
\label{fig:coul}
\end{figure}

The parameter
$\beta$ also considerably affects the Coulomb potential through
Eq.~(\ref{coul_eff}), as it is shown in Fig.~\ref{fig:coul}. We will see
in Sec.~\ref{data} that the effective potential can become very small
with respect to the retardation term.

\section{Comparison with other approaches}
\label{compar}

Since the relative time is introduced in the same way for our model and
the COQM, both approaches share some common properties. For instance,
the relative time part of our wave function
Eq.~(\ref{fo_ret}) and the counterpart in the COQM have the same
form. In this last model, for two quarks with the same mass $m$, we have
\cite{ishi86,ishi92}
\begin{equation}
A(\sigma)=\left(\frac{\kappa}{\pi}\right)^{1/4}
\exp\left(-\frac{\kappa}{2}\sigma^2\right) \quad \text{with} \quad
\kappa=\sqrt{\frac{mK}{2}},
\end{equation}
where $K$ is a constant related to the interquark potential
\begin{equation}\label{CO1}
U=-\frac{1}{2}K \bm r^{2}.
\end{equation}
Moreover this model predicts also linear Regge trajectories, with a
constant square mass shift $\Delta M^2_{\text{COQM}}$ due to the
retardation which is given by
\begin{equation}
\Delta M^2_{\text{COQM}} = - 2\sqrt{2mK}.
\end{equation}
Besides this formal similarities, the physical content of these two
models are nevertheless very different: A comparison is made in
Table~\ref{tab:comp}.

\begin{table}[h]
\caption{Comparison between the COQM and our model.}
\label{tab:comp}
\begin{ruledtabular}
\begin{tabular}{lll}
 & \bf COQM & \bf Our model \\
\hline
\bf Mass operator & square mass operator & ordinary Hamiltonian \\
\bf Confinement & quadratic & linear \\
\bf Allowed masses & non zero & all \\
\bf Treatment of retardation & exact & in perturbation \\
\bf Time and space decoupling & complete and exact & partial and
approximate
\\
\bf Shift in energy & negative and constant & negative and state
dependent
\\
\bf Relative time wave function & gaussian with constant size & gaussian
with state dependent size \\
\end{tabular}
\end{ruledtabular}
\end{table}

A more usual method to get relativistic
corrections (including retardation) is to consider the $v^{2}/c^{2}$
terms of the Bethe-Salpeter
equation or of the Wilson loop formulation of QCD. This was done for
example in Refs.~\cite{barc89,olss92}. In Ref.~\cite{olss92}, the
relativistic
correction to the linear confinement potential due to the retardation
is given by
\begin{equation}\label{corr2}
\Delta M=-\left\langle \frac{a}{m^{2}}\left(\frac{\ell(\ell+1)}{2r}+
rp^{2}_{r}\right)\right\rangle .
\end{equation}
As this correction term is
obtained by a expansion in $v^2/c^2$, one could expect that the
best agreement with our method will be obtained for large quark
masses. However, these corrections are very different from our
term~(\ref{omegahh}). In particular, the correction~(\ref{omegahh})
decreases with the
quantum numbers $n$ and $\ell$ because it decreases with the
constituent mass, but the contribution~(\ref{corr2})
increases with these quantum numbers. So, our approach does not lead to
a nonrelativistic limit compatible with previous works.

Actually, in order to obtain linear Regge trajectories, it is necessary
to obtain $\Delta M \propto 1/M$ (see Eq.~(\ref{m2meson})) for light
quark systems. But, for heavy quark systems, one could expect that
$\Delta M \propto 1/m^2$ with $m \approx M/2$. Our model clearly misses
this transition. A proper treatment in a covariant formalism with
constraints
could probably cure this flaw, but it is out of the scope of this paper.

\section{Comparison with experimental data}
\label{data}

We saw in Sec.~\ref{compar} that our formalism had unusual features,
compared with already known results. So it is important to verify if it
can correctly reproduce the experimental data. We will make here
such an attempt with the $n\bar{n}$ and $b\bar{b}$ mesons, in order to
check our model in different mass domains. The ingredients we put in our
``realistic" model are: The Hamiltonian~(\ref{muss})
including the string correction \cite{bada02,buis043}, the exact
retardation term $\Delta M$, the effective Coulomb
potential~(\ref{coul_eff}) treated as a perturbation and the quark
self-energy \cite{Sim1}.

The string correction has the following form
\begin{equation}\label{stringcor}
\Delta M_{\text{string}}=-\frac{a \ell (\ell +1)\langle 1/r \rangle}
{\langle \mu \rangle (6 \langle \mu \rangle + a \langle r \rangle)}.
\end{equation}
The quark self-energy is due to the color magnetic moment
of the quark propagating through the vacuum
background field, and it has been shown that it brings a contribution to
the meson mass given in the symmetrical case by
\begin{equation}\label{dHqses}
\Delta M_{\text{QSE}}=-\frac{fa}{\pi}
\frac{\eta(m/\delta)}{\langle \mu \rangle},
\end{equation}
with $f\in\left[3,4\right]$ and
$\delta\in\left[1.0,1.3\right]$~GeV. The function $\eta$ is given, for
instance, in Ref.~\cite{buis043}, in which a
more detailed discussion about the quark self-energy and its
consequences on the meson spectrum can be found.
Let us note that $\eta(0)=1$.

\begin{table}[h]
\caption{Our set of physical parameters.}
\label{tab:params}
\begin{ruledtabular}
\begin{tabular}{lcl}
$a=0.192$~GeV$^{2}$&\ & $\alpha_{S}=0.4$ \\
$m_{n}=0$ &\ & $\delta=1.0$~GeV \\
$m_{b}=4.660$~GeV &\ & $f=3.0$ \\
\ &\ & $$ $\lambda=1.0$~GeV \\
\end{tabular}
\end{ruledtabular}
\end{table}

The physical parameters we use are given in Table~\ref{tab:params}.
We have tried as much as possible to choose standard values:
$a=0.192$~GeV$^{2}$ and $\alpha_{S}=0.4$ are widely used, and
$m_{b}=4.660$~GeV is an acceptable value for the $b$ quark. The
parameter $f$ is fixed at $3$, which is the value obtained by
simulations in unquenched lattice QCD calculations \cite{DiGia1}.
The value $\delta=1.0$~GeV is used, but this choice has a very little
influence \cite{buis043}. Finally, we fix $\lambda=1.0$~GeV in order to
find, with the above parameters, the $n\bar n$ ground state near the
center of gravity of the $\pi$ and $\rho$ mesons, at 612.5~MeV (see
below).

Since our model includes neither the spin
($S$) nor the isospin ($I$) of the mesons, the experimental data chosen
are the spin and isospin averaged masses for the light mesons, denoted
$M_{\text{av}}$ (see Table~\ref{tab:results1}). These are given by
\cite{Brau98}
\begin{equation}
M_{\text{av}}=\frac{\sum_{I,J}(2I+1)(2J+1) M_{I,J}}{\sum_{I,J}(2I+1)(2J+
1)},
\end{equation}
with $\vec J=\vec L+\vec S$. $M_{I,J}$ are the different masses of the
states with the same orbital angular momentum $\ell$. For the $b\bar b$
mesons, we present results for the radial excitation of the $\Upsilon$
(see Table~\ref{tab:results2}). These
data are taken from Ref.~\cite{pdg}.

\begin{table}[htb]
\caption{Comparison between the spin averaged masses $M_{\text{av}}$ of
some $n\bar n$ family states (see Ref.~\cite{buis043} for more details)
and the numerically computed
masses $M$ (\ref{mtot}) of our model. The first three
columns present the different states used to compute the spin averaged
masses. The last column gives the contribution of the effective Coulomb
term.}
\label{tab:results1}
\begin{ruledtabular}
\begin{tabular}{lcccc}
Family & $N^{2S+1}L_{J}$ & $M_{\text{av}}$ (GeV) & $M$ (GeV) &
$\langle \tilde{V}_{C}\rangle$ (MeV)\\
\hline
$\rho$ & $1^{2S+1}S_{J}$ & $0.613\pm0.011$ & $0.631$ & $-19$ \\
$a_{2}(1320)$ & $1^{2S+1}P_{J}$ & $1.265\pm0.011$ & $1.235$ & $-6$ \\
$\rho(1700)$ & $1^{2S+1}D_{J}$ & $1.676\pm0.012$ & $1.669$& $-3$ \\
$a_{4}(2040)$ & $1^{2S+1}F_{J}$ & $2.015\pm0.012$ & $2.022$& $-1$ \\
\end{tabular}
\end{ruledtabular}
\end{table}

We see in Table~\ref{tab:results1} that our
results are in good agreement with the spin averaged masses. For all
states, the relative error is below 3\%. In each case, the
influence of the effective Coulomb term is very small and could be
neglected. Its role in lowering the mass is played by the contribution
of the retardation. Despite this unusual effect, our
approach allows to correctly reproduce the spin averaged masses.
With a smaller value for the parameter $\lambda$, the contribution of
the Coulomb term could be enhanced, but probably to values below those
obtained in other potential models.

As it can be seen in Table~\ref{tab:results2}, the relative error
between the data and our result is below 1\% for the mesons $\Upsilon$.
As expected in this case since $\beta$ is larger, the contribution of
the Coulomb potential is
larger. Heavy quark systems being more sensitive to the very
short range part of the interaction, better results could probably be
obtained by using a running coupling constant $\alpha_S(r)$. But this is
out of the scope of this paper.

\begin{table}[htb]
\caption{Same as in Table~\ref{tab:results1}, but for the masses
$M_{{\rm exp}}$ of some $b\bar{b}$
mesons. The experimental error bars are smaller than $10$~MeV and are
not indicated.}
\label{tab:results2}
\begin{ruledtabular}
\begin{tabular}{lccccc}
State &  $N^{2S+1}L_{J}$ &$M_{{\rm exp}}$ (GeV) & $M$ (GeV) &
$\langle \tilde{V}_{C}\rangle$ (MeV)\\
\hline
$\Upsilon(1S)$    & $1^{3}S_{1}$& 9.460 & 9.582 & $-87$ \\
$\Upsilon(2S)$    & $2^{3}S_{1}$ & 10.023& 9.990& $-52$ \\
$\Upsilon(3S)$    & $3^{3}S_{1}$& 10.355 & 10.294& $-40$\\
$\Upsilon(4S)$    & $4^{3}S_{1}$ & 10.580& 10.555& $-33$\\
$\Upsilon(10865)$ & $5^{3}S_{1}$ & 10.865 & 10.788& $-29$\\
$\Upsilon(11020)$ & $6^{3}S_{1}$ & 11.019& 11.002& $-26$\\
\end{tabular}
\end{ruledtabular}
\end{table}

\section{Conclusion}
\label{conclu}

In this paper, the retardation effects in mesons are taken into account
by the introduction of a non zero relative time in the rotating string
Hamiltonian \cite{dubi94,guba94}, following a procedure inspired by the
covariant oscillator quark model \cite{ishi86,ishi92}. Treated as a
perturbation, the part of the total Hamiltonian containing the
retardation terms is a harmonic oscillator in the relative time
variable, with an effective reduced mass and an effective restoring
force both depending on eigenstates of the Hamiltonian independent of
the relative time. The fundamental state of this oscillator gives the
contribution of the retardation to the masses as well as the relative
time part of the wave function. The introduction of the retardation also
affects the Coulomb part of the interaction, which is replaced by an
effective damped potential. Systems containing two particles with the
same mass are only considered, but our approach leads to several
interesting results.

In the light quark sector, the contribution of the retardation is not
negligible (around 200~MeV for massless quarks) but it is small enough
to justify a perturbative treatment. Within this approach, the Coulomb
interaction is strongly reduced but the meson masses are lowered by the
contribution of the retardation. The linearity of the Regge trajectories
is preserved, which is the most important feature of our model. At last,
the relative time wave function is a gaussian function centered around
zero, which confirms the validity of the equal time ansatz in first
approximation.

When the quark mass increases, the contribution of the retardation to
the meson masses decreases slowly. The relative time wave function
becomes more and more peaked around zero, as expected. Unfortunately,
our model does not lead to a nonrelativistic limit in agreement with
previous works. This is probably due to the treatment of the relative
time which is not compatible with a proper elimination of this
unphysical degree of freedom \cite{sazd86,crat87}.

Nevertheless, when the effective Coulomb potential and the quark
self-energy \cite{buis043} are included in our rotating string model
with the retardation effects, meson spectra can be computed in good
agreement with the experimental data.

This work must be considered as a trial to compute easily the
retardation effects in mesons. The simplified and approximate approach
used here has no firm theoretical basis, but it shows that the
contribution of these mechanisms to the masses could be non negligible.
Moreover, the importance of the retardation correction are not
controlled by the quark mass $m$ but by its constituent--state
dependent--mass $\langle \sqrt{\vec p\,^2+m^2} \rangle$ \cite{morg99}.

\appendix*

\section{Coefficients of the Lagrangian}\label{annexe}

The coefficients used in the Lagrangian~(\ref{step2}) are defined by
\begin{subequations}
\label{acdef}
\begin{eqnarray}
a_{1}&=&\mu_{1}+\mu_{2}+\int^{1}_{0} d\beta\, \nu , \label{a1def} \\
a_{2}&=&\mu_{1}-\zeta(\mu_{1}+\mu_{2})+\int^{1}_{0} d\beta \,
(\beta-\zeta)\, \nu , \label{a2def} \\
a_{3}&=&\mu_{1}(1-\zeta)^{2}+\mu_{2}\zeta^{2}+\int^{1}_{0} d\beta\,
(\beta-\zeta)^{2}\, \nu, \label{a3def} \\
a_{4}&=&\int^{1}_{0} d\beta\,
\left(\eta^{2}\nu-\frac{a^{2}}{\nu}\right),
\label{a4def} \\
c_{1}&=&\int^{1}_{0} d\beta \, \eta\,  \nu, \label{c1def} \\
c_{2}&=&\int^{1}_{0} d\beta\,  (\beta-\zeta)\,  \eta\,  \nu.
\label{c2def}
\end{eqnarray}
\end{subequations}
Using the extremal values~(\ref{etadef}) and (\ref{nu0}) of the
auxiliary fields $\eta$ and $\nu$, we can compute the following
relations
\begin{subequations}
\begin{eqnarray}
&&\int^{1}_{0}d\beta \nu=\frac{ar}{y_t}\left[\arcsin
s\right]^{y_{1}}_{-y_{2}}, \\
&&\int^{1}_{0}\frac{d\beta}{\nu} =\frac{1}{2ary_t}\left[s\sqrt{1-s^{2}}+
\arcsin s\right]^{y_{1}}_{-y_{2}}, \\
&&\int^{1}_{0}d\beta
(\beta-\zeta)\nu=-\frac{ar}{y_t^{2}}\left[\sqrt{1-s^{2}}\right]^{y_{1}}_
{-y_{2}},\\
&&\int^{1}_{0}d\beta
(\beta-\zeta)^{2}\nu=\frac{ar}{2y_t^{3}}\left[-s\sqrt{1-s^{2}}+\arcsin s
\right]^{y_{1}}_{-y_{2}}.
\end{eqnarray}
\end{subequations}

In the symmetrical case, we have $\mu_{1}=\mu_{2}=\mu$,
$m_{1}=m_{2}=m$, $y_{1}=y_{2}=y$, $y_t=2y$, $\zeta=1/2$, $\phi=1/2$, and
$\tilde{\mu}=\mu/2$. Equations~(\ref{acdef}) are in this case given by
\begin{subequations}\label{acdef3}
\begin{eqnarray}
a_{1}&=&2\mu+\frac{ar}{y}\arcsin y,\\
a_{2}&=&0,\\
a_{3}&=&\frac{\mu}{2}+\frac{ar}{8y^{3}}\left(-y\sqrt{1-y^{2}}+\arcsin y
\right), \\
a_{4}&=&-\frac{a}{2ry}\left(y\sqrt{1-y^{2}}+\arcsin y\right)+\kappa^{2}
\frac{ar}{8y^{3}}\left(-y\sqrt{1-y^{2}}+\arcsin y\right),\\
c_{1}&=&-\kappa\frac{ar}{2y^{2}}\sqrt{1-y^{2}},\\
c_{2}&=&\kappa\frac{ar}{8y^{3}}\left(-y\sqrt{1-y^{2}}+\arcsin y\right).
\end{eqnarray}
\end{subequations}

\end{document}